%% file: test.tex
\documentclass{article}
\usepackage{spconf,amsmath,graphicx}
\usepackage{fancyhdr}
 \usepackage{epsfig,amssymb,amsmath,multirow,boldline,graphicx,makecell,hyperref,subcaption,textcomp}
\usepackage{kotex}
\usepackage{float}
\usepackage{multirow}

\usepackage{xcolor}

\newcommand{\skm}[1]{\textcolor{red}}

\title{Domain Mismatch Robust Acoustic Scene Classification\\using Channel Information Conversion}

\name{Seongkyu Mun$^1$, Suwon Shon$^2$}
\address{Clova AI Research, Naver Corp. Seongnam, South Korea$^1$\\
MIT Computer Science and Artificial Intelligence Laboratory, Cambridge, MA, USA$^2$\\
{\small \tt sk.moon@navercorp.com \qquad swshon@csail.mit.edu }}%

\begin{document}
%\ninept
%
\maketitle
\begin{abstract}
In a recent acoustic scene classification (ASC) research field, training and test device channel mismatch have become an issue for the real world implementation. To address the issue, this paper proposes a channel domain conversion using factorized hierarchical variational autoencoder. Proposed method adapts both the source and target domain to a pre-defined specific domain. Unlike the conventional approach, the relationship between the target and source domain and information of each domain are not required in the adaptation process. Based on the experimental results using the IEEE  detection  and  classification of  acoustic  scenes  and  event 2018 task 1-B dataset and the baseline system, it is shown that the proposed approach can mitigate the channel mismatching issue of different recording devices.

\end{abstract}
\begin{keywords}
acoustic scene classification, factorized hierarchical variational autoencoder, domain adaptation
\end{keywords}
\section{Introduction}
\label{sec:intro}
%\sscm{you have a bunch of papers about event recognition in interspeech and ICASSP use that in the introduction to convince the reviewer!}
Acoustic Scene Classification (ASC) is a task that classifies input sounds into specific acoustic scenes, such as office, park, airport, tram, etc. In previous ASC researches, transfer learning~\cite{Mun_ICASSP, Mun_Inter}, attention mechanism~\cite{att1, att2} and DB augmentation~\cite{Mun2017dcase, gist} were proposed to improve ASC performance.
Recently, researches on the ASC have been intensively studied on the IEEE Detection and Classification of Acoustic Scenes and Events (DCASE) 2016-2018 challenges~\cite{Mesaros2018ieeetrans,Mesaros2018iwaenc,Mesaros2018dcase}. Due to the simple and clear task of classifying pre-defined scene labels for sound data of specific length, various techniques from acoustic signal processing fields, such as speaker recognition and music information retriever, have been tried. All of the top teams in the last three years used Convolutional Neural Network (CNN)-based architectures and additionally used I-vectors~\cite{Eghbal2016}, Generative Adversarial Networks (GAN) based DB augmentation~\cite{Mun2017dcase} and harmonic-percussive source separation based pre-processing~\cite{Sakashita2018,Han2017}, respectively. 
Unlike 2016 and 2017, the 2018 DCASE challenge task 1 added subtask B, which addresses the dataset recorded with different devices~\cite{Mesaros2018dcase}. In real-world environments, the device mismatching issue is inevitable, so the new subtask has a practically important issue. The task consists of a relatively large source domain A (recorded by device A) and a relatively small target domain B and C (recorded by device B and C, respectively). Submissions of the challenge task 1-B are ranked by classification accuracy of target devices B and C. To the best of our knowledge, during the challenge period, there were no submitted technical reports that handle different device issue directly. After the challenge, in order to address the related issue, a paper using GAN based domain adaptation has been submitted for the following DCASE 2018 workshop~\cite{Gharib2018}. The adaptation module (from ‘target’ to ‘source’ domain) and domain discriminator are optimized through adversarial training, and the ASC performance was improved on the DCASE 2018 task 1-B DB. 

Although the aforementioned approach effectively adapted the device domain, there is a limitation that DB of the target domain is required for training the adaption module. This limitation could be critical in some cases. For example, when input sounds are received via other unseen devices or web-streaming, it is difficult to gather sufficient target domain DB for training the domain adaptation module. 

\begin{figure}[ht]
     \centering
     \includegraphics[width=\linewidth]{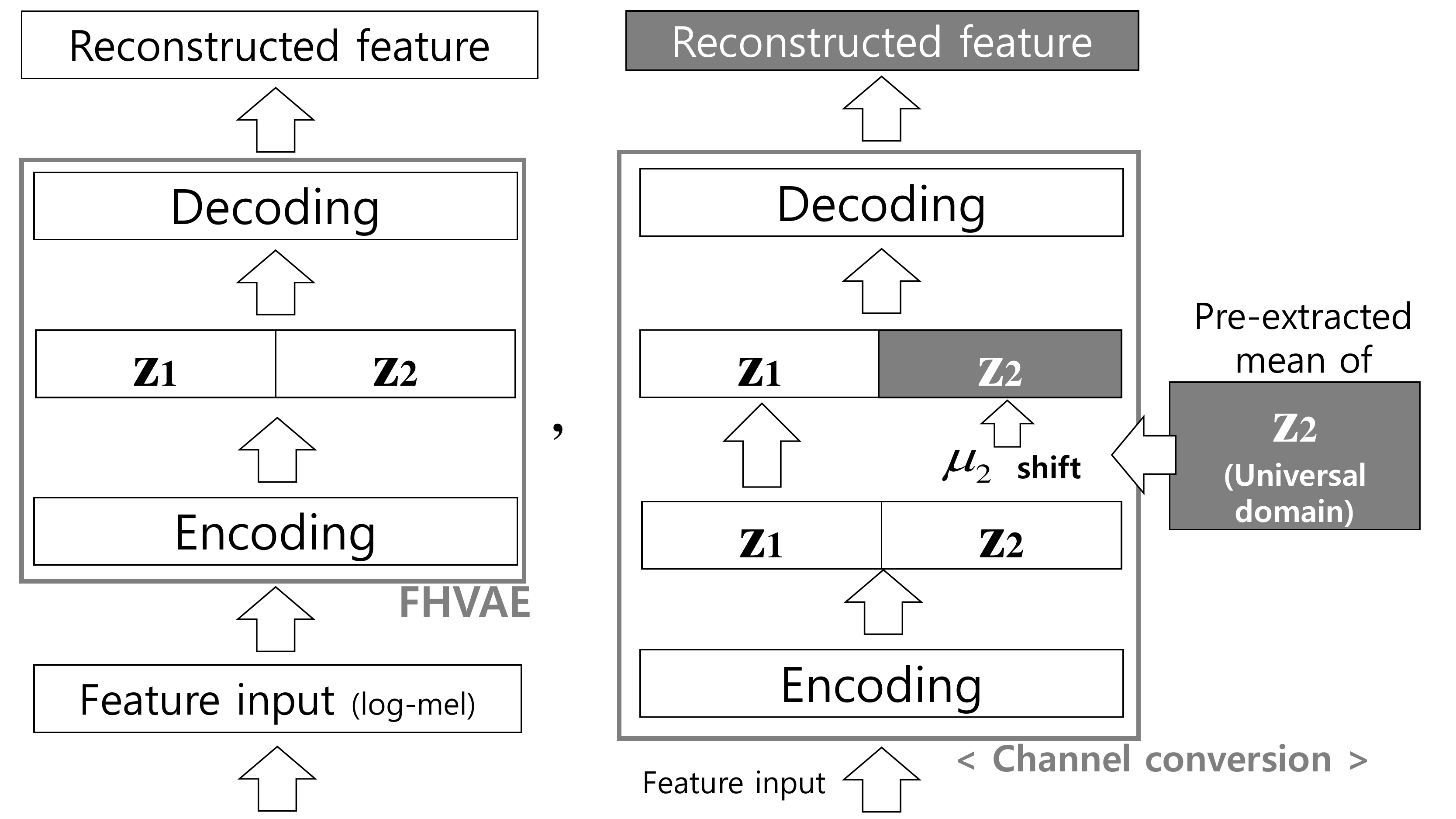}
     \caption{The structure of FHVAE (Left) and concept of channel conversion (Right)}
    \label{fig:fhvae_concept}
\end{figure}

To address the issue, this paper proposes an adaptation from \emph{target or source} to \emph{other specific} domain, not target to source domain or vice versa. Proposed method adapts the source domain as well as the target domain. As shown in Figure~\ref{fig:fhvae_concept}, the device domain (channel) related component ($z_2$) is disentangled from the input signal, then it is shifted to the other domain (e.g. universal domain), and the reconstruction process is followed. Since $z_2$ components of the input features are mapped to the specific domain, the relationship between the target and source domain and information of each domain are not required in the adaptation process. In order to implement the aforementioned process, we utilized Factorized Hierarchical Variational AutoEncoder (FHVAE)~\cite{Hsu2017Nips}, which shows notable performance improvement in voice conversion and sequential information representation~\cite{Hsu2017interspeech,Hsu2018icassp, Hsu2018inter_scalable}. We adapted input features by using FHVAE for generating factors of the acoustic scene and device-related component and mapping the device related component to the other specific domain. Based on the experimental results using the DCASE 2018 task 1-B dataset and the baseline system, it is shown that the proposed approach can mitigate the channel mismatching issue of different recording devices.

\section{Acoustic Scene classification in Channel mismatched condition}
\label{sec:format}
\subsection{Channel related component disentanglement using FHVAE}
\begin{figure}[ht]
     \centering
     \includegraphics[width=40mm]{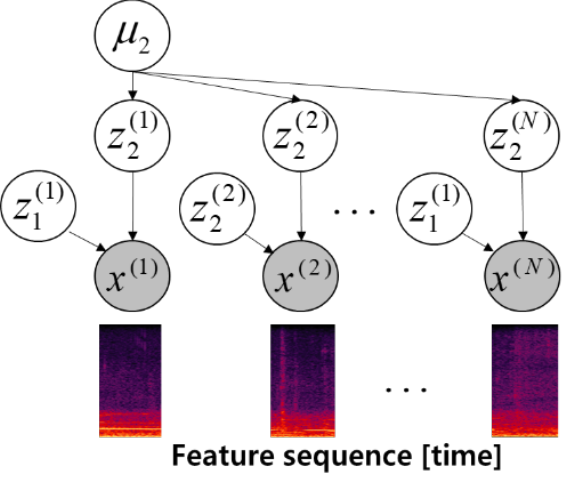}
     \caption{Graphical illustration of the FHVAE generative model. Grey nodes denotes observed variables, and white nodes are the latent variables}
    \label{fig:model}
\end{figure}

In this section, we briefly describe the FHVAE for our ASC system. More details of FHVAE can be found in~\cite{Hsu2017Nips,Hsu2017interspeech,Hsu2018icassp}. The FHVAE~\cite{Hsu2017Nips} is a variant of variational autoencoder~\cite{Kingma2014iclr} that models a probabilistic hierarchical generative process of sequential data, and learns disentangled and interpretable representations. Generation of a sequence of $N$ segments  involves one sequence-level latent variable, $\mu_2$ , and $N$ pairs of  segment-level latent variable  $z1$ and $z2$. $\mu_2$, the prior component of sequence-dependent is drawn from $p(\mu_2)=N(\mu_2|0,\sigma^2_{\mu_2}I)$. $N$ i.i.d latent segment variables $Z_1=\{z^{(n)}_1\}^N_{n=1}$ are drawn from a global prior $p(z_1)=N(z_1|0,\sigma^2_{z_1}I)$. $N$ i.i.d. latent sequence variables $Z_2=\{z^{(n)}_2\}^N_{n=1}$ are drawn from a sequence-dependent prior $p(z_2|\mu_2)=N(z_2|\mu_2,\sigma^2_{z_2}I)$. At last, $N$ i.i.d. sub-sequences $X = \{ x^{(n)} \}_{n=1}^N$ are drawn from $p(x | z_1, z_2) = \mathcal{N}(x | f_{\mu_x}(z_1, z_2), diag(f_{\sigma^2_x}(z_1, z_2)))$, where $f_{\mu_x}(\cdot,\cdot)$ and $f_{\sigma^2_x}(\cdot,\cdot)$ are parameterized by a decoder neural network.~\cite{Hsu2018inter_scalable,Shon2018unsuper}.
Since the exact posterior inference is intractable, FHVAEs introduce an inference model to approximate the true posterior. In this work, we followed the latest training method of the FHVAE research~\cite{Hsu2018inter_scalable}. Figure~\ref{fig:model} shows the aforementioned model.

\begin{figure}[ht]
     \centering
     \includegraphics[width=\linewidth]{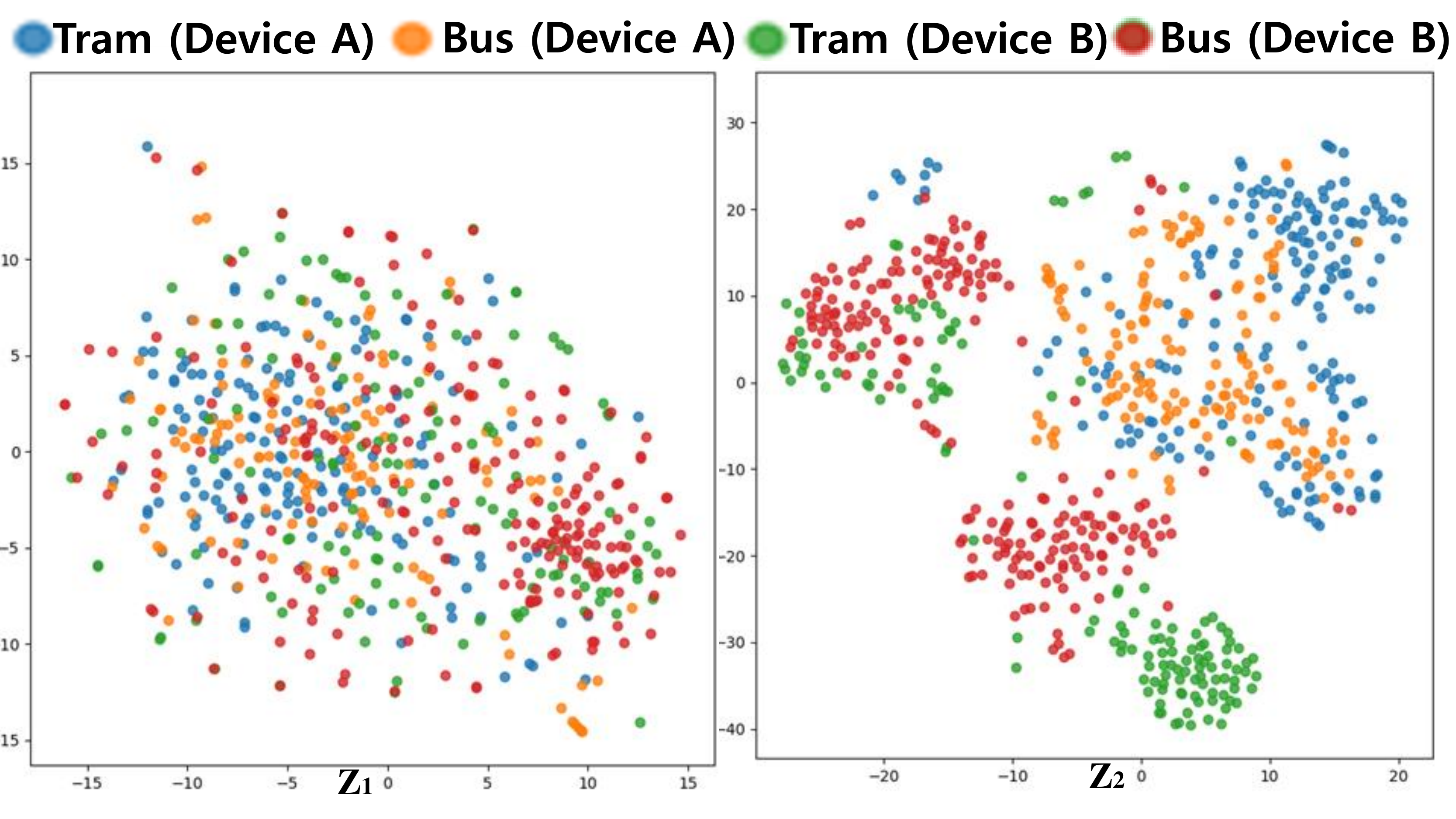}
     \caption{Scatter plots of t-SNE projected $z_1$ and $z_2$ with models trained on DCASE 2018 task 1-B}
    \label{fig:scatter}
\end{figure}

By imposing a sequence-dependent prior to $z_2$, the model is encouraged to represent with $z_2$ the generating factors that are relatively consistent within a sequence. For example, such factors can include microphone frequency response and room impulse response. On the other hand, $z_1$ tends to encode information about the residual generating factors that change from segment to segment, such as acoustic scene related audio events. In order to compare the characteristics of the two latent variables, 2-dimension $t$-Distributed Stochastic Neighbor Embedding (t-SNE) projection examples of $z_1$ and $z_2$ distributions are shown in Figure~\ref{fig:scatter}. Each point represents one segment. The FHVAE model was trained using DCASE 2018 task 1-B development train set DB. Note that the device label information was not used in the FHVAE training process. Detail configurations of the model and inference will be discussed in section 3. As shown in Figure~\ref{fig:scatter}, $z_1$ shows channel invariant characteristic compared to $z_2$, and $z_2$ distribution has a tendency to be clustered by each channel subset. The result of $z_2$ plot shows that blue dots (Tram, Device A) are closer to orange dots (Bus, Device A) of the same recording device, compared to the green dot (Tram, Device B) of the same acoustic scene class.

Assuming that the latent variable $z_2$ contains channel discriminative information, we propose a process of shifting $z_2$ values for equalizing (adapting) channel components. In addition, based on previous researches of channel invariant feature representation~\cite{Shon2018unsuper, Hsu2018inter_unsuper}, we conducted an experiment of using $z_2$ as a feature input of the classifier (without reconstruction) for performance comparison.

\subsection{Channel conversion using latent variable shifting }
\begin{figure}[ht]
     \centering
     \includegraphics[width=\linewidth]{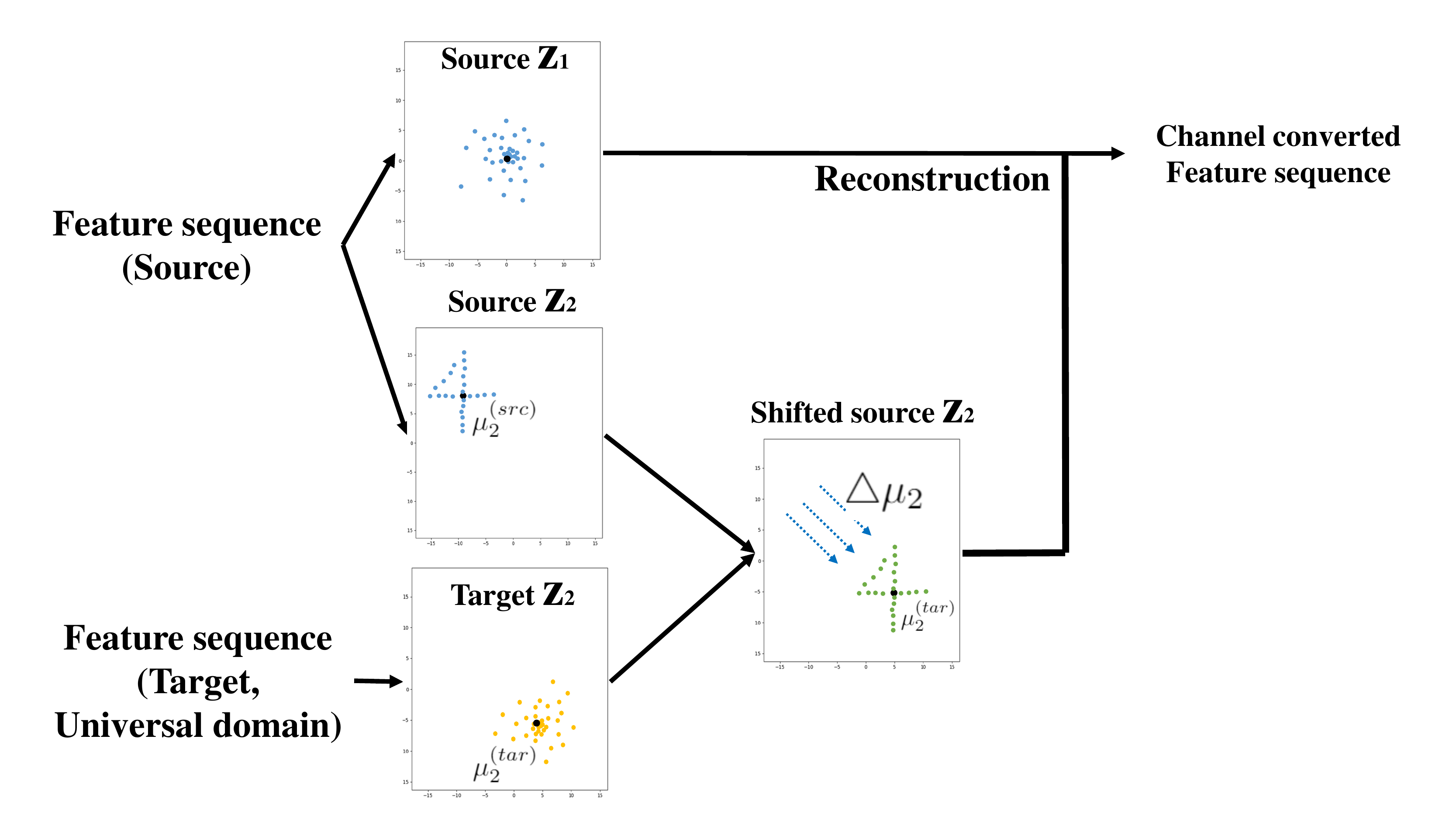}
     \caption{Concept example of mean shifting for channel conversion}
    \label{fig:channel_shift}
\end{figure}

To utilize latent variable $z_2$ to convert channel, we obtained motivation from the previous research of~\cite{Hsu2017Nips}. For transforming sequence-level attributes while preserving segment-level attributes, we conducted the mean $\mu_2$ shifting process. Based on the FHVAE framework, channel conversion is equivalent to mapping the distribution of latent sequence variables of the source sound $X^{(src)}$ to target sound $X^{(tar)}$. For each segment in $X^{(src)}$, we shift $Z_2^{(src,n)}$ by modifying $\mu_2$,($\bigtriangleup\mu_2=\mu_2^{(tar)}-\mu_2^{(src)}$) while keeping $Z_1^{(src,n)}$ unaltered, as shown in Figure~\ref{fig:channel_shift}. Based on the method, we shift the $z_2$ values using the pre-obtained $\mu_2^{(tar)}$ (general mean of target domain DB) from the target domain, as shown in Figure~\ref{fig:fhvae_concept}.

\begin{figure}[ht]
     \centering
     \includegraphics[width=\linewidth]{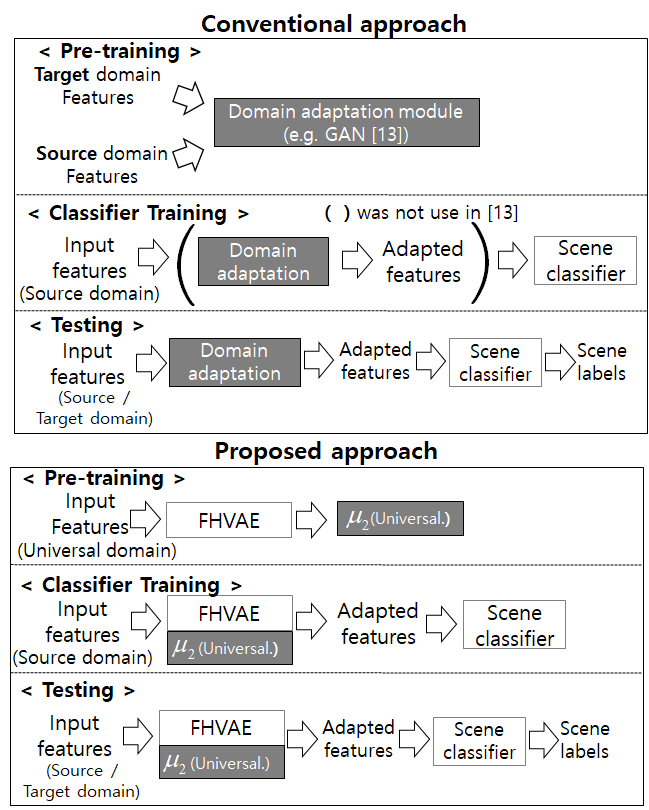}
     \caption{Comparison of domain adaptation process between conventional and proposed approach}
    \label{fig:proposedconcept}
\end{figure}

Here, as in the conventional domain adaptation~\cite{Gharib2018}, the target domain can be set to device A for channel conversion. (Device B, C to Device A) However, as mentioned above, input sounds might be unseen target domain input in the real-world environment, so modeling the relationship between specific domains cannot be a general solution for domain adaptation. Therefore, we propose an approach of converting unspecified domains into a target domain, not the approach of converting specific domains into a target domain. Following the proposed approach, the source domain (e.g. development-train set in DCASE 2018 task 1-B), which is generally a large size DB, also have to be converted. For the convenience of notation, we set the ‘universal domain’ to the our proposed target domain. As shown in Figure~\ref{fig:proposedconcept}, we trained the FHVAE and obtained $\mu_2$ of the universal domain through pre-training step. Using the $\mu_2$ from the universal domain, channel conversion is conducted on the training DB and then the acoustic scene classifier is trained using the converted DB as shown in Figure~\ref{fig:proposedconcept}. Since the channel converted DB is used for both training and testing step, the classifier is not affected by channel mismatching. Compare to the conventional method, the DB from the target domain is not required in the pre-training step for channel conversion.

\subsection{Acoustic scene classifier}
For acoustic scene classification, we used the DCASE 2018 task 1 baseline system~\cite{Mesaros2018dcase} to concentrate more on the performance changes of domain adaptation rather than classification performance itself. The baseline system implements a CNN based approach, where log mel scale-band energies are extracted for each 10-second signal, and a network consisting of two CNN layers and one fully connected layer is trained to assign scene labels to the audio signals. Detail configuration of baseline system can be found in the official website~\footnote{http://dcase.community/challenge2018/task-acoustic-scene-classification}.
\input{perform_table.tex}
%\begin{figure}[ht]
%     \centering
%     \includegraphics[width=\linewidth]{configure.pdf}
%     \caption{DCASE 2018 Task 1-B DB configuration [3]     %\label{fig:configure}
%\end{figure}

\section{Experimental settings and results}

%Task 1 DB configuration is shown in Figure~\ref{fig:configure}.
For the acoustic scene classification experiment, DCASE 2018 task 1 dataset was used.  The dataset was recorded in six large European cities, in different locations for each scene class. For each recording location, there are 5-6 minutes of audio. The original recordings were split into segments with a length of 10 seconds with one label from the pre-defined scene labels (Airport, Indoor shopping mall, Metro station, Pedestrian street, Public square, Street, Tram, Bus, Metro, and Urban park). 
The total size of the device A dataset is 8640 segments of 10 seconds length, i.e. 24 hours of audio. The dataset contains 2 hours of parallel data recorded with all three devices (A, B and C). The amount of data is as follows:\\
\\
• Device A: 24 hours (8640 segments, 864 segments / scene)\\
• Device B: 2 hours (720 segments, 72 segments / scene)\\
• Device C: 2 hours (720 segments, 72 segments / scene)\\
\\
The training subset contains 6122 segments from device A, 540 segments from device B, and 540 segments from device C. The test subset contains 2518 segments from device A, 180 segments from device B, and 180 segments from device C. Since the DB label set of the final evaluation DB set, which was used for challenge ranking, were not released, the training and testing were conducted with the DB configuration used during the challenge. We follow configurations of the FHVAE model in~\cite{Hsu2018inter_scalable}, but the input feature dimension is modified to the number of dimensions in the DCASE baseline ASC model. (log-mel of 40 dimensions) DB augmentation with SNR 10 and 15 dB level was conducted with white Gaussian noise in the FHVAE training. (3-times larger than original DB volume) We did not use augmented DB for classifier training. The experimental results are shown in Table 1.

For comparison with the baseline system, we first used $z_1$ latent variables, which showed channel invariant characteristics, as input features without reconstruction. System performance was slightly improved on device B and C, but performance for device A was worsened. This is because FHVAE extracts information about the residual generating factors that change from segment to segment for $z_1$, rather than scene class discriminative information. On the other hand, since $z_2$ variables may have scene information, it is necessary to reconstruct the original scene information through the decoding process after converting the channel component as in the proposed method. The performance evaluation of the proposed method consists of three DB configurations. All the cases improved performance for device B and C, but performance was slightly decreased on device A set. First, similar to~\cite{Gharib2018}, we set device A as a target domain and obtaining $\mu_2^{(tar)}$ for training and testing ASC system. Since the approach uses device A domain information, performance of device A set was not decreased much compared to baseline, but performance improvement of device B and C domain was limited. When device B and C domain information in development-train set were used in FHVAE, the highest ASC performance for device B, C was achieved. This is an obvious result because the system used the information of the target device in the pre-training step. In the last case, we obtained $\mu_2^{(tar)}$ using DB from DCASE 2016 and 2017 task 1 development set~\cite{Mesaros2018ieeetrans,Mesaros2018iwaenc}. Although the DB configuration and labels of DCASE 2016 and 2017 are different from the 2018 set, we could use these DBs, because the FHVAE system is based on unsupervised training. In this case, even though the FHVAE does not utilize domain information of the training and test DB of the scene classifier, a higher performance, compared to the case of using the device A set, was achieved using the various acoustic scene DBs. The results show the effectiveness of the proposed method, and it is noteworthy that the domain adaptation can be conducted only by using the universal domain information without the target or source domain information.
Unlike the baseline model of DCASE 2018 task 1, a baseline model used in the previous domain adaptation research ~\cite{Gharib2018} more intensively trained on the device A domain rather than device B or C. Therefore, it is difficult to conduct a proper comparison experiment with the proposed methods. For comparison, it is necessary to conduct adversarial adaptation on the DCASE 2018 task 1 model~\cite{Mesaros2018dcase} in the future work.

\section{Conclusion and Futurework}

This paper proposes FHVAE based channel conversion for ASC in channel mismatching condition. Proposed latent variable shifting method shows performance improvement on the DCASE 2018 task 1-B DB. Especially, the proposed approach shows ASC performance improvement in mismatched condition by using only universal scene DB in pre-processing, without source domain (device A) and target domain (device B, C) information.
For the future work, we plan to research for speech enhancement in a similar approach. In the situation of source domain (clean speech) and the target domain (unseen noisy speech), it would be meaningful if noisy channel adaptation is possible with only universal speech DBs, regardless of the target noise type.
\section*{Acknowledgements}
\small
The authors would like to thank Donmoon Lee and Dr. Yoonchang Han for valuable discussion to inspire ideas of the paper.
% \vfill\pagebreak
\clearpage

% \section{REFERENCES}
% \label{sec:refs}

\bibliographystyle{IEEEbib}
\bibliography{refs}

\end{document}

%% file: perform_table.tex
\begin{table}[]
\centering
\resizebox{0.5\textwidth}{!}{%
\begin{tabular}{l|c|c|c}
\hlineB{2}
\multirow{2}{*}{Average accuracy on scene classes (\%)}
 & Device A & Device B,C & \multirow{2}{*}{Averaged} \\
 & (source) & (target) \\ \hlineB{2}
DCASE 2018 baseline ~\cite{Mesaros2018dcase}& 58.9& 45.6& 52.3\\ \hline
Domain adapted DCASE Kaggle model ~\cite{Gharib2018} & \textbf{65.3} & 31.7 & 48.5\\  \hlineB{2}
\textbf{Proposed Methods}\\ \hline
Using $Z_1$ as feature & \multirow{2}{*}{56.3} & \multirow{2}{*}{46.8} & \multirow{2}{*}{51.6} \\
(without reconstruction) & & \\ \hline
Shift $z_2$ to $\mu_2^{(tar)}$ of device A & 58.2 & 47.1 & 52.7 \\ \hline
Shift $z_2$ to $\mu_2^{(tar)}$ of device B and C & 57.8 & \textbf{51.2} & \textbf{54.5}\\ \hline

Shift $z_2$ to $\mu_2^{(tar)}$ of external scene DB & \multirow{2}{*}{57.6} & \multirow{2}{*}{49.8} & \multirow{2}{*}{53.7} \\
(universal acoustic scene domain) & & \\ \hline

%Shift $z_2$ to $\mu_2^{(tar)}$ of external scene DB\\ (Universal domain) & 57.6 & 49.8 & 53.7 \\ \hline
%[8] with [17] baseline non adapted  & 65.3 & 20.3 & \\ \hline
\hlineB{2}

\end{tabular}
}
\caption{Class average accuracy (\%) on DCASE 2018 task 1-B development dataset. Note that the challenge was ranked by average accuracy on device B, C}
\end{table}